\documentclass[aip,amsmath,amssymb,reprint]{revtex4-1}
\usepackage{graphicx}
\usepackage{dcolumn}
\usepackage{bm}
\usepackage[utf8]{inputenc}
\usepackage[T1]{fontenc}
\usepackage{mathptmx}

\begin{document}

\preprint{AIP/123-QED}

\title{Stable Polarization Entanglement based Quantum Key Distribution over Metropolitan Fibre Network}
\author{Yicheng Shi}\email{cqtsy@nus.edu.sg}
\author{Soe Moe Thar}
\author{Hou Shun Poh}
\author{James A. Grieve}
\affiliation{Centre for Quantum Technologies, 3 Science Drive 2, National University of Singapore, 117543 Singapore}

\author{Christian Kurtsiefer}\email{phyck@nus.edu.sg}
\author{Alexander Ling}
\affiliation{Centre for Quantum Technologies, 3 Science Drive 2, National University of Singapore, 117543 Singapore}
\affiliation{Department of Physics, National University of Singapore, Blk S12, 2 Science Drive 3, 117551 Singapore}

\date{\today}

\begin{abstract}
We demonstrate a quantum key distribution implementation over deployed dark telecom fibers with polarisation-entangled photons generated at the O-band. One of the photons in the pairs are propagated through 10\,km of deployed fiber while the others are detected locally. Polarisation drifts experienced by the photons propagating through the fibers are compensated with liquid crystal variable retarders. This ensures continuous and stable QKD operation with an average QBER of 6.4\% and a final key rate of 109 bits/s.
\end{abstract}

\maketitle

\section{Introduction}

Quantum Key Distribution (QKD) enables two users to share a common encryption key that is secret to any third parties. Early QKD protocols such as BB84~\cite{BB84} were "prepare-and-measure" schemes, with practical derivatives such as SARG04~\cite{SARG04} and decoy states~\cite{Lo2005Decoy}. This was complemented by the invention of entanglement-based protocols such as E91~\cite{PhysRevLett.67.661} and BBM92~\cite{PhysRevLett.68.557}, with quantitative extensions through device-independent QKD~\cite{Acin2005Devindep}. Both types of QKD protocols have been proven theoretically secure and have been studied extensively over the decades~\cite{Meyrs96, LoChau1999, ShorPreskill00, 10.1007/978-3-540-30576-7_21}.

For prepare-and-measure protocols, a trusted random number generator is
required to provide randomness in the state preparation process. This is not
required for entanglement-based QKD protocols, where randomness of the key
originates from the measurement process itself. Entanglement-based QKD also
does not rely on a true single photon source or a decoy state mechanism to mitigate a photon number splitting attack, and has fewer possible side
channels than typical prepare-send scenarios. As such, entanglement based QKD is less vulnerable to attacks in practical implementations~\cite{PhysRevA.61.052304}.

Both freespace and optical fibre links have been used as the transmission channel for distributing entangled photon pairs~\cite{PhysRevLett.84.4729}. Due to low optical attenuation in the atmosphere, the channel loss over freespace links can be as low as 0.07\,dB/km at high altitudes~\cite{PhysRevLett.98.010504}. For protocols using polarisation entanglement, the state of the photons is well preserved during freespace transmission. Early implementations of freespace QKD used optical telescopes to send and receive photons over a range~\cite{Rarity2001, Hughes_2002, Kurtsiefer02}, reaching over hundred of kilometers~\cite{PhysRevLett.98.010504}. Further more, this range can be extended to thousands of kilometers by utilizing satellites as intermediate nodes~\cite{Liao2017}. 

Optical fiber links, on the other hand, are suitable when a line-of-sight is not available. Fiber-based QKD generally operates over shorter range (<100\,km) due to optical attenuation of light in the fiber. This is, however, enough to cover metropolitan areas where a fiber network is available~\cite{PhysRevX.8.021009, Poppe:14, Dynes2019, joshi2019trustednodefree}. 

The available telecom single mode fiber conforms to the ITU G.652 standard~\cite{ITU652}. To maximize range, fiber-based QKD systems can use entangled photons generated at telecom C-band (1530-1565\,nm) where fiber absorption is at its minimum (0.2\,dB/km)~\cite{Treiber_2009, Wengerowsky6684, Wengerowsky2018, joshi2019trustednodefree}. The O-band in (1260-1360\,nm) is another choice of wavelength, with an abosrption loss of about 0.32\,dB/km. The total loss over fiber transmission in a realistic link is always higher due to the presence of splicing and patching points.

The presence of dispersion effects is another possible limiting factor to the
performance of entanglement-based QKD over fiber. Entangled photon pairs are
usually generated via a Spontaneous Parametric Down Conversion (SPDC) process,
which leads to photons of relatively large bandwidth when performed in
nonlinear optical crystals, compared to photons generated with lasers. Such
wideband photons experience then significant chromatic dispersion in the fiber ($\sim$ 18\,ps/nm/km at 1550\,nm)~\cite{smf28}. This increases the uncertainty in timing correlation between the entangled photons, leading to a lower signal to noise ratio, eventually reducing the final key rate. The effect of chromatic dispersion can be migitated by using dispersion-shifted fiber~\cite{Treiber_2009}, or by using entangled photons at telecom O-band operating on either side of the zero-dispersion wavelength of the fiber~\cite{doi:10.1063/1.5088830}. 

For QKD protocols using polarisation encoding, an optical fibre cannot be simply regarded as a pure loss channel. When propagating through the fibre, an arbitrary rotation is applied to the polarization state of photons and causes basis mismatch. In addition, fiber Polarisation Mode Dispersion (PMD) can cause degradation of polarisation entanglement for broadband photons~\cite{PhysRevA.63.012309,Brodsky:11}. Both effects increase the Quantum Bit Error Rate (QBER),  reducing the rate of key generation. While the polarisation rotation can be compensated~\cite{Xavier_2009}, the presence of polarisation mode dispersion has led to a preference of time-bin encoding over polarisation encoding in fiber-based QKD implementations~\cite{PhysRevA.63.012309, Fasel2004,doi:10.1063/1.5089784}. However in recent years, manufacturers are able to make single mode telecom fibers with much lower PMD value ($\leq0.04\,\text{ps}/\sqrt{\text{km}}$)~\cite{smf28}, which makes polarisation encoding possible even for relatively broadband entangled photons.

In this work we report an entanglement-based QKD system implemented over a 10\,km deployed fibre in a metropolitan area. The setup uses the BBM92 protocol~\cite{PhysRevLett.68.557}, with polarization entangled photon pairs generated at telecom O-band to minimize the effect of chromatic dispersion. The polarisation rotation due to the fiber is compensated using liquid crystal variable retarders (LCVRs) after which the polarisation state is stable during several hours of continuous QKD operation. 

\section{Implementation}
\begin{figure}
\includegraphics[width=0.95\linewidth]{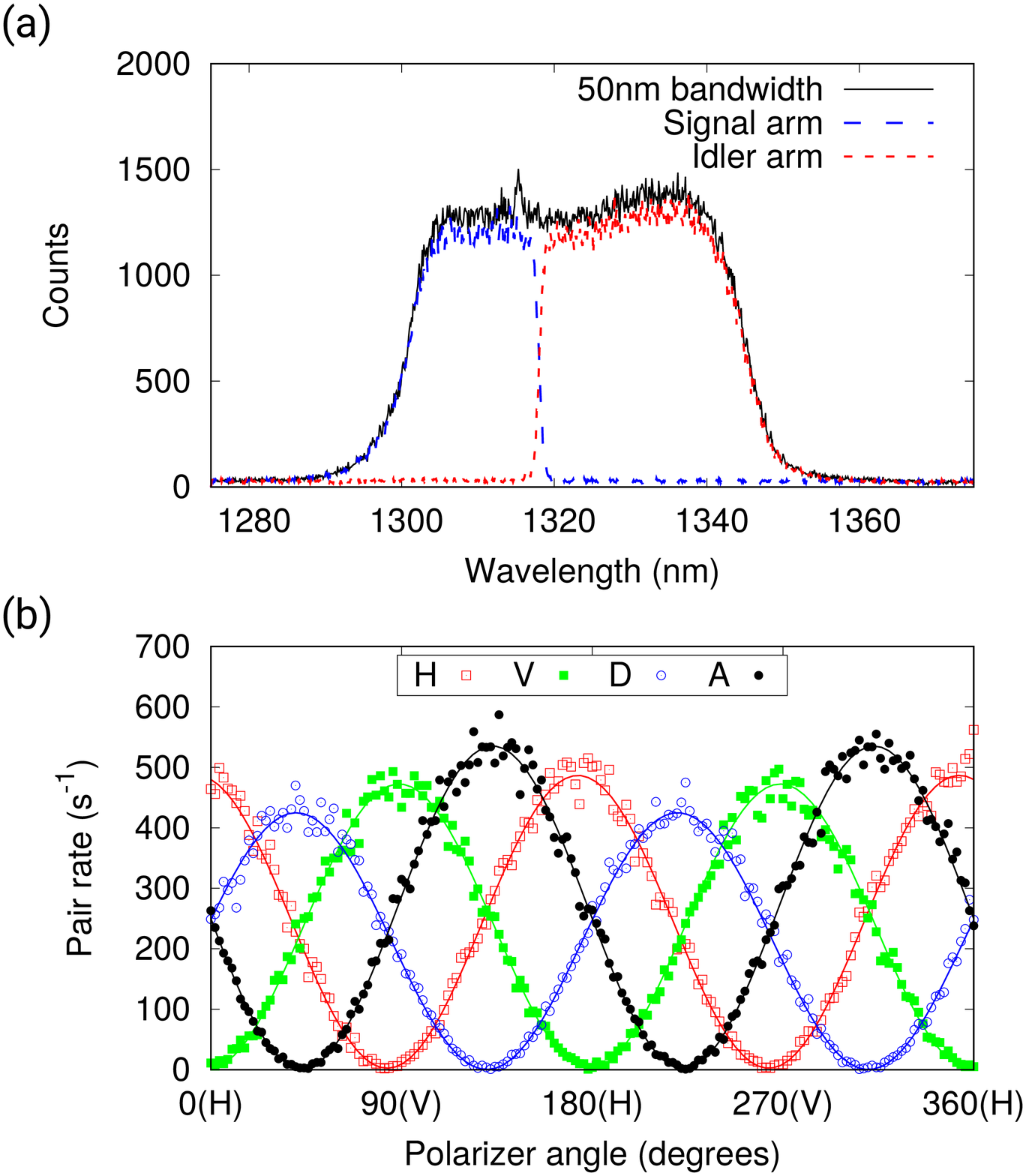}
\caption{\label{fig:source_stuff}(a) Spectrum of the Typr-0 SPDC photons. The black trace shows the 50 nm bandwidth defined by the bandpass filter applied. Signal (blue) and idler (red) photons are separated using a wavelength division demultiplexer. (b) Polarization correlation in both H/V and D/A bases measured at the entanglement source.}
\end{figure}

\begin{figure}
\includegraphics[width=0.95\linewidth]{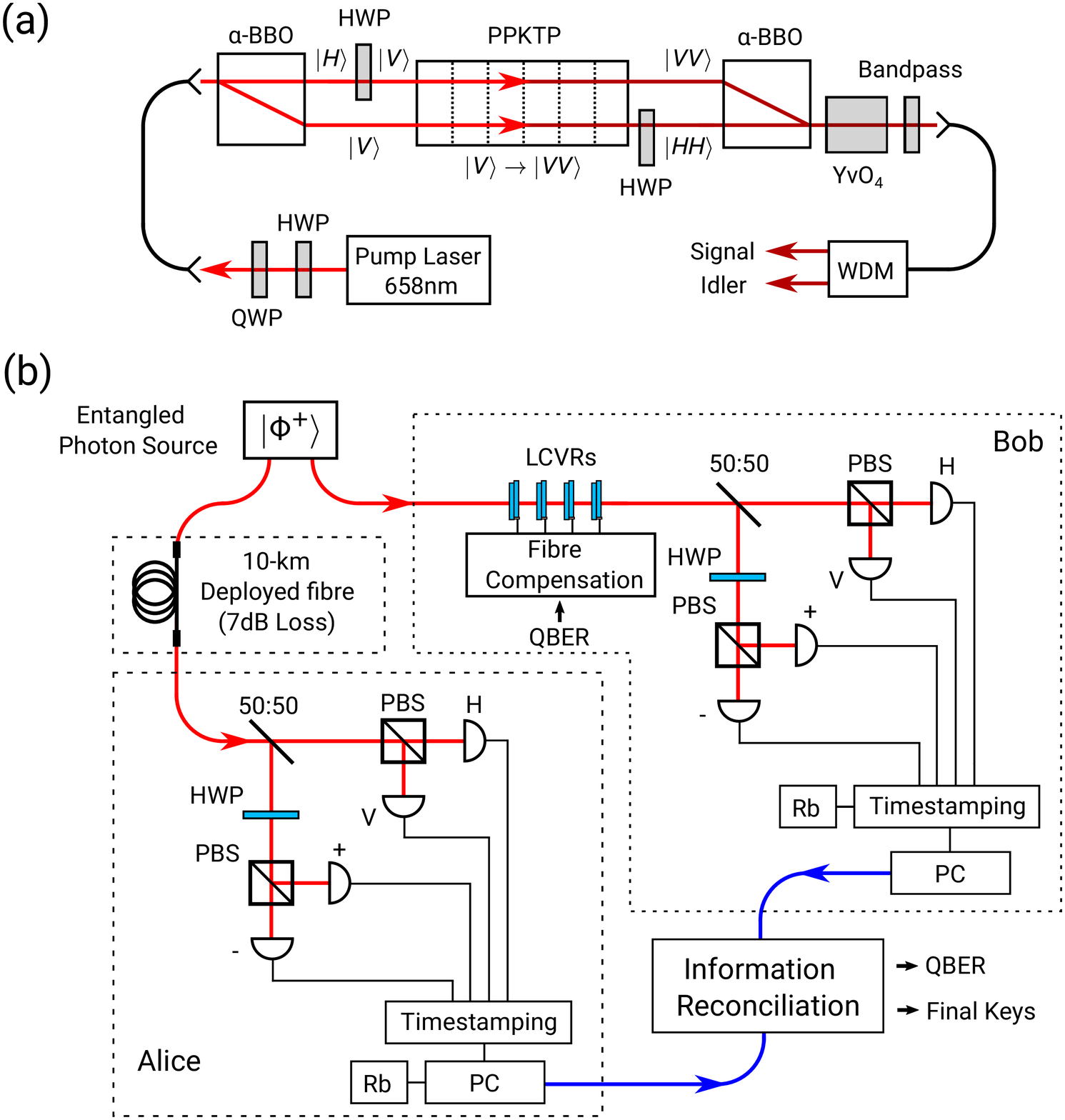}
\caption{\label{fig:setup} (a) Experimental schematic of the entanglement source. The pump photons are split to two paths and undergo type-0 SPDC inside the PPKTP crystal. The polarization state in the lower path is rotated 90 degrees by a half-wave plate. The two paths are then recombined to create a $|\Phi^+\rangle$ state. (b) QKD setup over 10\,km deployed fiber link. The fiber loops back to the lab to simplify the experimental procedure. The Alice and Bob nodes are run on independent clocks. Alice's analyzer is connected to the entanglement source via the 10\,km deployed fiber while Bob's setup is locally connected using a short patch cord. The two hosting PCs are connected to the same local area network in order to exchange timestamp data for coincidence identification.}
\end{figure}

The entangled photon pairs are generated via type-0 SPDC inside a periodically-poled potassium titanyl phosphate (PPKTP) crystal, shown in Fig.~\ref{fig:setup} (a). The crystal is pumped by a grating stabilized laser diode at 658\,nm, emitting photon pairs that are degenerate at 1316 nm. The bandwidth of the downconverted photons is limited to about 50 nm by a bandpass filter. As shown in Fig.~\ref{fig:source_stuff} (a), a wavelength division demultiplexer with an edge at approximately 1316 nm is used to separate the signal and idler photons. The entanglement state is prepared by placing the PPKTP crystal inside a linear beam-displacement interferometer~\cite{doi:10.1063/1.5124416}. The photon pairs are prepared in a state:
\begin{align*}
|\Phi^+\rangle=\frac{1}{\sqrt{2}}(|H_A H_B\rangle + |V_A V_B \rangle)
\end{align*}
with polarisation visibility over 98\% in both horizontal/vertial (H/V) and diagonal/anti-diagonal (D/A) bases (Fig.~\ref{fig:source_stuff} (b)). With the pump power of 2.4 mW, we observed a local pair rate of $4300$\,s$^{-1}$.

The setup of the QKD system is shown in Fig.~\ref{fig:setup} (b). The entangled photon pairs are distributed to two nodes, Alice and Bob, with a polarisation analyzer placed on each side. A 10\,km telecom fiber connects the source to Alice's analyzer while Bob's setup is connected locally via a short patchcord. 

\begin{figure}
\includegraphics[angle=-90,width=0.95\columnwidth]{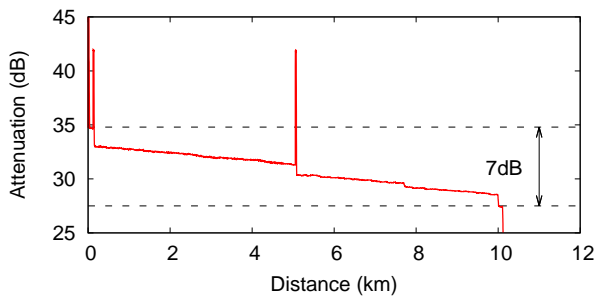}
\caption{\label{fig:fiber_stuff} Optical time-domaine reflectometer trace of the deployed fibre, identifying a high reflection loss point about 5 km away from both end points. Two more points with high reflection/absorption loss are also identified about 100 meters from the end points, which is due to a 100 meter patching cable between the deployed fiber and the laboratory setup.}
\end{figure}

The 10\,km telecom fiber is deployed underground by Singapore Telecommunications Limited in a loop configuration with both ends located at Center for Quantum Technologies, National University of Singapore. Measurement using an optical time-domain reflectometer (OTDR) shows a total fiber length of 10.4\,km with about -7\,dB channel loss (Fig.~\ref{fig:fiber_stuff}). The optical absorption of the fiber contributes only about -4\,dB to the total channel loss, with another -3\,dB loss due to reflections at patching points and losses at splicing points. The total PMD of the fiber link is about 0.1\,ps, which is smaller than the coherence time of the signal and idler photons ($\sim$0.23\,ps for 25\,nm bandwidth at 1310\,nm).

Polarization change due to fiber is compensated by placing a set of 4 LCVRs
before Bob's analyzer setup. This compensation only needs to be applied to one
of the photons from each pair to restore the initial $|\Phi^+\rangle$ state
after fiber propagation. In some implementations with optical fibers on the
surface~\cite{Xavier_2009}, polarisation compensation needs to be constantly
performed due to the rapid change of the polarization change, which severely
limits the operation continuouity of QKD.

The polarization stability of 10\,km deployed fibre in our setup was characterized by
sending in light with well-defined polarization state across the fiber and
monitoring the change in polarization with a
polarimeter~\cite{doi:10.1080/09500340600674242}. We find that the output
polarization drift was slow, with a typical 24\,h period associated with
day-night temperature change. Once the polarization rotation is compensated,
the fiber allows several hours of stable QKD operation even without running any active compensation scheme.

Upon receiving the photons, Alice and Bob follow the BBM92 protocol by
measuring polarizations in one of the two bases: H/V and
D/A~\cite{PhysRevLett.68.557}. The random detection basis choice is made by a non-polarizing beam splitter in each setup which transmits and reflects photons with equal probability~\cite{Rarity1994}. Four commercial Indium Gallium Arsenide Avalanche Photodiodes (InGaAs APDs) are used in each analyzer setup for single photon detection. The APDs diodes are cooled down to below -40\,$^\circ$C and are operated in freerunning mode with a nominal detection efficiency around 10\% and an average dark count rate of about 12000\,s$^{-1}$.  On each side, detected photons are timetagged to a resolution of 125 ps with a 4-channel timestamping device locked to a rubidium frequency standard~\cite{doi:10.1063/1.2348775}.

Recorded timestamp traces are continuously exchanged through a network connection between two hosting lab computers. To enable coincidence identification, the clocks on both sides are synchronized in advance by exploiting the intrinsic timing correlations of the SPDC photons~\cite{Ho_2009}. The uncertainty in the coincidence time difference is about 1.9\,ns (FWHM) due to fiber chromatic dispersion, detector timing jitter and other noise in the system~\cite{doi:10.1063/1.5088830}. For coincidence identification, a coincidence window of 0.5\,ns was chosen to optimize the coincidence/accidental ratio without losing too many coincidence events.

Raw key data are generated after coincidence identification and key sifting
following a typical BBM92 protocol. Error correction are then performed on
each block of raw key data accumulated over 25\,seconds~\cite{doi:10.1063/1.2348775} using a modified
CASCADE/BICONF algorithm~\cite{Brassard94}, following largely~\cite{Sugimoto00}. An estimated QBER is also obtained during error correction and is used to determine the amount of secure key bits to be extracted from the raw key bits. Privacy amplification is then performed on both sides for obtaining the final secure keys~\cite{doi:10.1137/0217014}.

We estimate a total system loss of -33$\,$dB in our entire QKD system, with
-7$\,$dB contributed by the total channel loss of the deployed fiber, -6$\,$dB
from the optical coupling loss in the polarisation compensation and analyzer
setup, and another -20\,dB solely due to the detection efficiency of the
InGaAs APDs on both sides. Therefore, detector efficiency is the dominant
contribution to the overall system loss in our setup. 

\section{Performance}

\begin{figure}[b]
\includegraphics[width=0.65\linewidth, angle=-90]{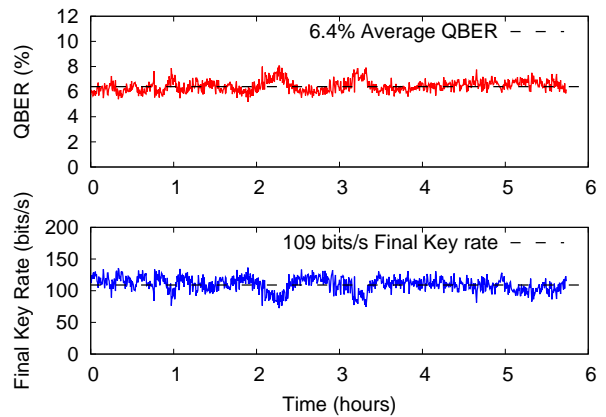}
\caption{\label{fig:log} QBER (top) and finaly key rate (bottom) logged over 5.7\,hours of continuous operation. Error correction and privacy amplification are performed over blocks of raw key bits integrated over 25 seconds. Data collection stopped after 5.7\,hours due to a detector failure.}
\end{figure}

With the 10\,km deployed fiber connected, the rate of detected single photons
is 40\,699\,s$^{-1}$ on Alice's analyzer, and 242\,125\,s$^{-1}$ at Bob's
side, respectively. We observe a coincidence rate of 670\,s$^{-1}$ and an accidental coincidence rate of 19\,s$^{-1}$. After an initial fiber compensation, the QKD setup operated continuously over 5.7\,hours until one of the detectors ceased operation due to a temperature overrun. The average sifted key rate after basis reconciliation is 340\,s$^{-1}$ with an average estimated QBER of 6.3\%. The final key rate after error correction and privacy amplification is about 109\,bits/second (Fig.~\ref{fig:log}).

Our final key rate is comparable to other reported entanglement-based
implementations at telecom C-band~\cite{joshi2019trustednodefree}, or at wavelengths
detectable by Silicon APDs~\cite{Poppe:04}. Secure transfer of messages with
this key rate is practical using one-time pad encryption for low bandwidth
communications such as command \& control of industrial systems. Alternatively,
the key can be utilized in fast encryption schemes using e.g. AES-256, with a
much more frequent re-keying compared to conventional methods~\cite{Dynes2019}. The key rate in
our demonstration is mainly limited by the low detection efficiency
($\sim$10\%) and high dark count rate ($\sim 10^4$\,s$^{-1}$) of the InGaAs
APDs in the setup. Significant increase in key rate is expected when replacing
them with superconducting nanowire detectors ($\sim$80\% detection
efficiency)~\cite{doi:10.1063/1.4896045}. As practical advantage of photons at
O-band, QKD can operate along the normal internet traffic with all channels in
C-band concurrently over the same fiber link.

\section{Conclusion}
We have demonstrated a stable entanglement-based quantum key distribution
system operating over a deployed telecom fiber of 10\,km distance following
the BBM92 protocol. Polarization-entangled photon pairs in the telecom O-band
minimize the effect of chromatic dispersion. The polarisation change in the fiber
due to fiber geometry and birefrincence is compensated with liquid crystal
variable retarders, enabling stable transmission of photon polarisation
states.
We operated the systems continuously for 5.7\,hours with an average QBER of
6.4\% and a final key rate of 109\,bits/s. The key rate performance is mainly
limited by the detection efficiencies and high dark count rate of the InGaAs
photodetectors.

\begin{acknowledgments}
This  research  was  supported  by  the  National  Research Foundation, Prime Minister’s Office, Singapore under its CorporateLaboratory@University Scheme, National University of Singapore, and Singapore Telecommunications Ltd.
\end{acknowledgments}


%

\end{document}